# Histogram-Driven Amplitude Embedding for Qubit-Efficient Quantum Image Compression

[1]Sahil Tomar,[2]Sandeep Kumar
[1,2] Central Research Laboratory, BEL, Ghaziabad, India

## ABSTRACT

This work introduces a compact and hardware-efficient method for compressing color images using near-term quantum devices. The approach segments the image into fixed-size blocks ("bixels") and computes the total intensity within each block. A global histogram with B bins is then constructed from these block intensities, and the normalized square roots of the bin counts are encoded as amplitudes into an n-qubit quantum state, where n=[$\log_2 B$]. Amplitude embedding is performed using PennyLane and executed on real IBM Quantum hardware. The resulting state is measured to reconstruct the histogram, enabling approximate recovery of block intensities and full-image reassembly. The method maintains a constant qubit requirement based solely on the number of histogram bins, independent of the image's resolution. By adjusting B, users can control the trade-off between fidelity and resource usage. Empirical results demonstrate high-quality reconstructions using as few as 5–7 qubits, significantly outperforming conventional pixel-level encodings in terms of qubit efficiency and validating the method's practicality for current NISQ-era quantum systems.

## 1. INTRODUCTION

With the explosive growth of high-resolution imaging and real-time streaming, classical codecs such as JPEG and HEIC built on transforms like the Discrete Cosine Transform are increasingly strained by modern demands on storage, bandwidth, and end-to-end security. Recent surveys reveal significant variability in classical-to-quantum mapping techniques, both in compression fidelity and circuit complexity, reinforcing the need for practical quantum encoding frameworks [1]. Histogram-based quantization, a classical strategy for tone-mapping and data summarization, has yet to be deeply explored in the quantum domain, despite offering a highly interpretable and compact way to approximate dense image statistics.

Quantum computing, through superposition, entanglement, and statistical encoding, introduces a fundamentally new paradigm for data representation, with the potential for compact encoding and novel processing capabilities for high-dimensional images [2], [3]. Foundational quantum image representations have made key strides but remain limited. The Flexible Representation for Quantum Images (FRQI) encodes grayscale intensity through rotation angles and positional qubits but is constrained to square, single channel images and introduces circuit overhead [2]. The Novel Enhanced Quantum Representation (NEQR) improves fidelity using binary registers but retains fixed-square assumptions and raises qubit requirements [4]. RGB extensions like NCQI and MCQI expand color capability but scale poorly due to complex preparation circuits and increased qubit counts [5].

Amplitude embedding presents an alternative by mapping normalized pixel values directly onto the amplitudes of an n-qubit state, achieving a logarithmic qubit count of $\log_2(N)$ for N pixels. While this approach is highly qubit-efficient, it often incurs deep circuit decompositions, suffers from limited robustness to noise, and lacks integrated compression workflows [1], [6], [7]. Histogram quantization and weighted redistribution, foundational in classical image equalization and block-level representation, are adapted here for quantum-compatible encoding. These techniques allow global summary statistics to be embedded with reduced dimensionality while preserving local structure [8]. Histogram binning is a fundamental technique in image processing where the continuous range of pixel intensity values is partitioned into a fixed number of discrete intervals, known as bins. Each bin accumulates the number of pixels whose intensity falls within its corresponding range, effectively transforming the image's raw pixel data into a summarized distribution of tonal values [9].

The proposed method leverages a classical statistical technique, histogram binning to achieve efficient quantum data representation. Histogram binning [9] not only reduces data dimensionality but also yields a compact and interpretable representation of an image's tonal distribution. In classical applications, histogram-based techniques underpin a wide array of tasks: contrast enhancement, through histogram equalization and its adaptive variants [8], noise reduction, via smoothing of intensity histograms to suppress minor fluctuations [10], and image segmentation, by identifying thresholds based on histogram valleys, particularly in multimodal distributions [11]. These capabilities arise from the histogram's ability to abstract spatial detail into a concise statistical form, supporting both interpretability and computational efficiency.

The proposed work brings classical histogram quantization into the quantum realm. An arbitrary H×W×3 RGB image is first partitioned into equal-sized blocks ("bixels") and compute each block's total intensity. These block sums are aggregated into a global histogram with B bins, where B controls the compression resolution and the normalized square root of the bin counts the amplitudes of an n-qubit quantum state, with n=$\lceil \log_2 B \rceil$. Using amplitude embedding on real IBM Quantum hardware, the state is measured to recover the histogram distribution and then redistribute each block's intensity according to its original intra-block weight pattern. This deterministic, no-training pipeline requires only n qubits, independent of image size and naturally balances fidelity against resource use through the choice of B.

This paper introduces a quantum framework that handles arbitrary-size, full-color images with a fully deterministic, end to end compression and reconstruction workflow by combining amplitude embedding, histogram summarization, per-block normalization into a unified scheme. The proposed work has been evaluated by comparing reconstructed image against the original image using standard fidelity metrics, mean squared error (MSE) and peak signal-to-noise ratio (PSNR) to demonstrate visually faithful recovery with significantly fewer qubits on real IBM Quantum hardware. The key novelties of the proposed approach include:

- Bixel-Based Encoding: The image is segmented into bixels (rectangular blocks), enabling scalable statistical abstraction at varying granularities. Each block is summarized using its intensity sum, which captures essential regional characteristics.
- Histogram-Driven Compression: A global histogram of block sums is constructed, and only this distribution is amplitude-embedded into a quantum state. This replaces the need for direct pixel-wise encoding and leads to dimensionality reduction.
- Qubit Usage Independent of Image Size: The number of qubits in the proposed approach depends solely on the number of histogram bins B, allowing compression of very large images using as few as 5 qubits.
- End-to-End Workflow: The method supports a complete quantum pipeline from classical preprocessing and histogram generation to amplitude embedding and reconstruction based on quantum measurements, with local redistribution guided by per-block weights.
- Benchmark-Driven Evaluation: The technique is evaluated by comparing reconstructed image against the original image using standard fidelity metrics such as MSE and PSNR, highlighting competitive performance at significantly reduced quantum resource cost.

## 2. PROPOSED METHODOLOGY

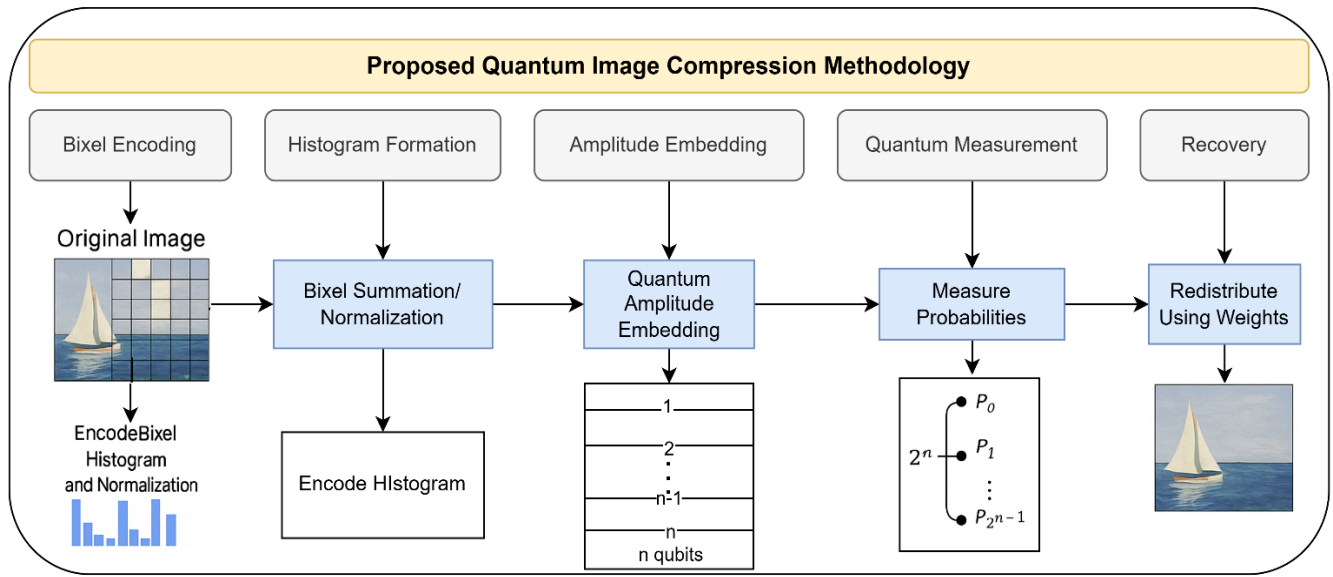

Figure 1: Overview of the proposed quantum image compression pipeline using bixel histogram encoding and amplitude embedding.

The compression pipeline is built upon a structured abstraction of the image using statistical summarization over small spatial regions. This process is divided into three stages: bixel encoding, histogram formation, and measurement-based recovery. Each step serves a distinct role in transforming the original image into a quantum-representable format with significantly reduced dimensionality and qubit requirements.

## 2.1 Bixel Encoding

The process begins by ingesting a classical RGB image of arbitrary resolution, denoted as $I \in R^{H \times W \times 3}$. All pixel values are normalized to the interval $[0, 1]$. The image is then segmented into non-overlapping rectangular patches called bixels, each of size $bixel_H \times bixel_W$. This segmentation facilitates localized summarization and ensures compatibility with histogram binning. To align with the bixel grid, zero-padding is applied to the image if H or W is not a multiple of the respective bixel dimension. Each bixel is treated as a flattened vector of length:

$$M = bixel_H * bixel_W * 3 \tag{1}$$

Total intensity is computed for each bixel:

$$S_b = \sum_{i=1}^{M} p_i \tag{2}$$

where $p_i$ is the ith pixel in the flattened bixel in eq. (2). To retain intra-block detail, a normalized weight vector is also stored for the bixel:

$$w_{b,i} = \frac{p_i}{S_b}, \text{with } \sum_{i=1}^{M} w_{b,i} = 1 \tag{3}$$

These weights from eq. (3) allow the original structure of the block to be approximately reconstructed later using a single scalar estimate.

## 2.2 Histogram Formation

The second stage aggregates all bixel intensity sums $\{S_b\}$ into a global histogram. This step serves as a quantization mechanism, reducing the diversity of block-level information into a smaller number of representative classes. A histogram of bin B is defined, where each bin covers a range of block sum values. The count of bixels falling into each bin forms a histogram vector $h \in R^B$. To convert this histogram into a quantum-ready format, it is first normalized into an amplitude vector:

$$a_k = \sqrt{\frac{h_k}{\sum_{j=1}^{B} h_j}} \tag{4}$$

Ensuring the normalization condition for amplitude encoding is maintained using eq. (4):

$$\sum_{k=1}^{B} a_k^2 = 1 \tag{5}$$

Since quantum circuits require vectors of length $2^n$ the number of qubits required is determined by:

$$n = \lceil \log_2 B \rceil \tag{6}$$

and zero-pad to reach length $2^n$. This vector is now ready to be embedded into an n-qubit quantum state.

## 2.3 Measurement and Recovery

The padded amplitude vector defines a normalized quantum state of the form:

$$|\psi\rangle = \sum_{k=0}^{2^n-1} a_k |k\rangle \tag{7}$$

Using AmplitudeEmbedding, this state is initialized. A full projective measurement is then performed on a computational basis, producing a probability distribution:

$$P_k = |\langle k | \psi \rangle|^2 \tag{8}$$

These probabilities approximate the original histogram counts and are scaled back to estimate the reconstructed block sums. For each bixel, a histogram bin is determined to which it originally belonged to (if a sum falls exactly on a bin edge, assign it to the lower-indexed bin), and the center of that bin is taken as its reconstructed sum $S'_b$. Using the previously stored weight vector $w_{b,i}$ from eq. (3), each pixel value is reconstructed in the block as:

$$r_i = w_{b,i} * S'_b \tag{9}$$

After reconstructing all bixels, they are reshaped and combined to form the final image. Any previously added padding is removed to restore the image to its original dimensions. This methodology enables a scalable compression mechanism where the number of required qubits is independent of the full image resolution. Instead, it depends only on the number of histogram bins B, which the user can tune to balance fidelity and quantum resource constraints.

## 3. EXPERIMENTAL SETUP AND EVALUATION METRICS

To validate the generality and robustness of the proposed quantum image compression pipeline, experiments were conducted on a "Standard" test images [12] of widely used test images in the image processing literature. These include the Lenna, Peppers, Cameraman, Mandrill and various other images, each exhibiting diverse textures, edges, and color distributions. Each image was evaluated to assess performance across different spatial granularities.

In addition to these benchmarks, the method was tested on a high-resolution real-world RGB image with dimensions 3403 × 5266 pixels, representing a complex natural scene. This scenario served as a stress test for scalability with respect to both image size and compression behavior. The framework supports arbitrary user-provided images, dynamically handling file input, automatic normalization, and padding to ensure alignment with the bixel grid.

For the core experimental setup, images were partitioned into non-overlapping bixels of size 32 × 32 pixels. The sum of pixel intensities in each block was computed, and a global histogram of all block sums was constructed using B=32 bins. This histogram was square root normalized and amplitude-encoded into a compact quantum state using n=⌈$\log_2 B$⌉=5 qubits. The embedding and measurement steps were executed on actual IBM Quantum hardware using the ibm_sherbrooke backend via Qiskit Runtime Service, with 4096 measurement shots per experiment. This ensured realistic assessment under noisy intermediate-scale quantum conditions.

The measured probability distribution was scaled to reconstruct the histogram and used to estimate each block's sum, which was then redistributed across pixels using the originally computed intra-block weight vectors. This approach preserved the relative structure within each block while encoding only the global histogram into the quantum state. To evaluate the fidelity of the reconstructed images, two standard metrics were used. MSE is computed as:

$$\text{MSE} = \frac{1}{H*W*3} \sum_{i=1}^{H} \sum_{j=1}^{W} \sum_{c \in \{R,G,B\}} (I_{ijc} - \hat{I}_{ijc})^2 \tag{10}$$

where $I_{ijc}$ and $\hat{I}_{ijc}$ denote the original and reconstructed pixel intensities at spatial coordinate (i,j) and color channel c. PSNR is then computed in decibels as

$$\text{PSNR} = 10 \log_{10}\left(\frac{\text{MAX}^2}{\text{MSE}}\right), \ \text{MAX} = 1.0 \tag{11}$$

Higher PSNR indicates closer fidelity to the original image. Runtime is also monitored for state preparation and measurement to characterize simulation overhead as a function of qubit count. Together, these settings and metrics provide a comprehensive view of both reconstruction quality and computational cost, enabling direct comparison with future variational or hardware-based implementations.

## 4. RESULTS AND DISCUSSION

The proposed method was first implemented on various images on a "Standard" test images [12] as shown in Figure 2. to assess performance across different spatial granularities. Finally, a random test image of high-resolution (3403×5266×3) is taken to be executed on a simulator for various bin sizes and then the same random test image is run on actual IBM hardware for 32 bin size.

Figure 3. illustrates how reconstruction quality and computational cost scale for a high-resolution test image with qubit count based on different bin values of histogram when executing the proposed amplitude-based encoding on the simulator. As the number of qubits n increases corresponding directly to the number of histogram bins, both the MSE and PSNR improve significantly. At 3 qubits, MSE remains relatively high (≈ 0.026) and PSNR is low (≈ 15 dB), indicating poor reconstruction. By 4 qubits, fidelity improves markedly, with MSE dropping to around 0.007 and PSNR rising to approximately 25 dB. The most dramatic gains occur between 5 and 7 qubits. At 5 qubits, the MSE falls to ~0.002 while PSNR reaches about 38 dB; at 6 qubits, the MSE dips below 0.001 and PSNR exceeds 50 dB. By 7 qubits, MSE is reduced to under 0.0003, and PSNR climbs beyond 60 dB, with runtimes still under half a second. Beyond this point, although PSNR continues to increase, exceeding 75 dB by 8 qubits and runtime begins to rise gradually as circuit complexity and state preparation overhead increase. This trend demonstrates that even with a modest number of qubits, the proposed approach achieves high-fidelity image reconstruction. The super linear improvement in PSNR per added qubit highlights the power of amplitude-only encoding for compact yet expressive quantum image representation.

To place the amplitude-based encoding strategy within the context of real quantum hardware, Table 1 compares three established image encoding formats: FRQI, NEQR, and NCQI with the histogram-based approach. Whereas FRQI and

NEQR require ⌈$\log_2 N$⌉+1 qubits (grayscale angles) or ⌈$\log_2 N$⌉+8 qubits (8-bit grayscale registers), and NCQI requires ⌈$\log_2 N$⌉+2 qubits for RGB, In contrast, the histogram method encodes a bixel-level intensity distribution using only ⌈$\log_2 B$⌉ qubits, regardless of N, and imposes no strict aspect-ratio constraints beyond simple zero-padding to align block boundaries.

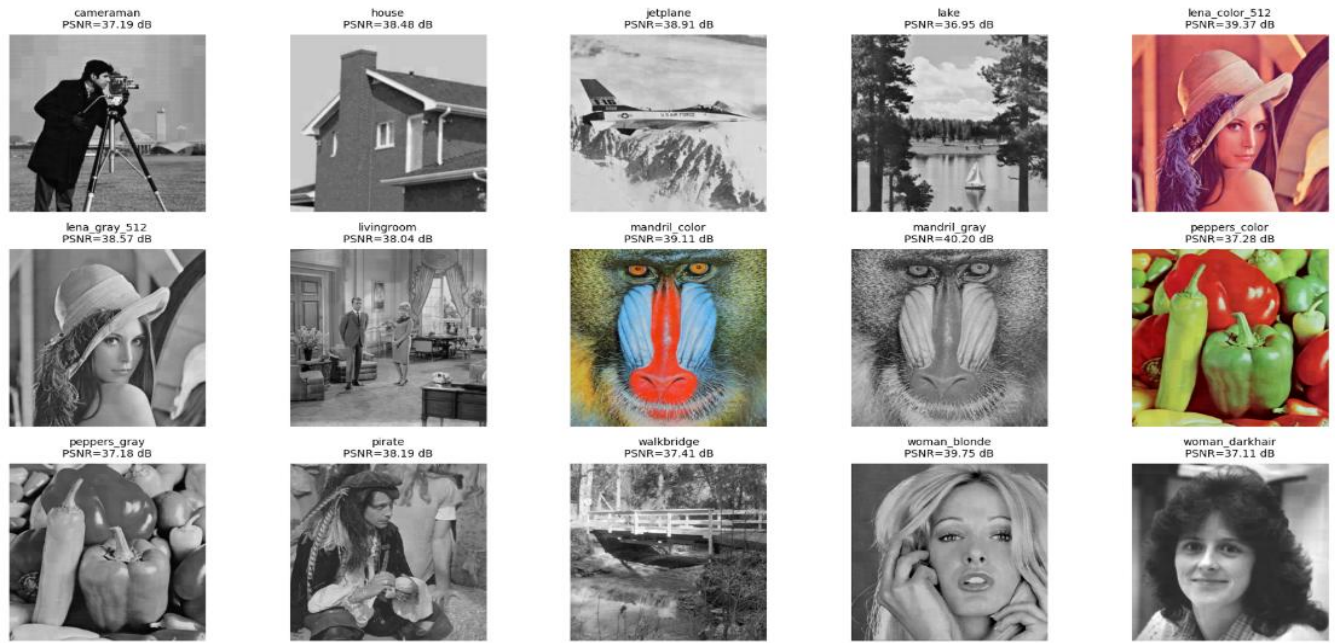

Figure 2: Reconstructed outputs of 15 standard grayscale and color test images using the proposed 4-qubit histogram-based quantum compression framework. All images were segmented into 32×32 bixels. The PSNR values shown above indicate high-fidelity reconstruction using amplitude embedding on a simulated quantum backend. Notably, the proposed method achieves consistent performance across diverse image types (natural, textured, synthetic, and portraits) using a constant number of qubits, independent of image resolution

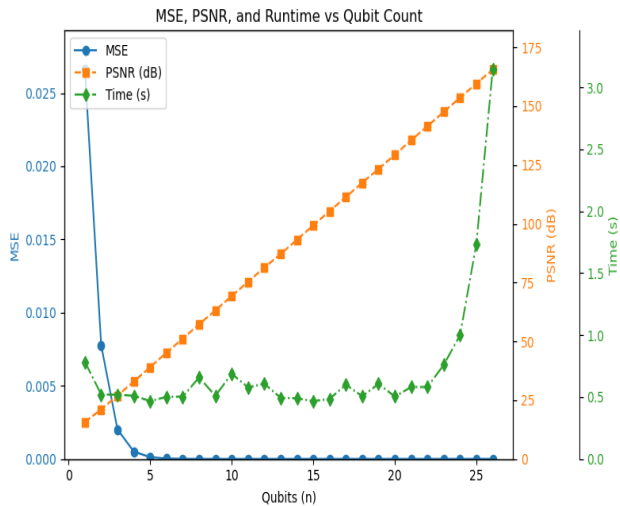

Figure 3: MSE, PSNR, and Runtime vs. Qubit Count for Histogram-Based Quantum Image Encoding on simulator for a random test image.

In the hardware implementation on IBM's ibm_sherbrooke device, an input image is first partitioned into 32×32 blocks, each block's total intensity is summed, and these sums populate a histogram of B bins. A square root normalized version of this histogram is padded to length $2^{\lceil \log_2 B \rceil}$ and loaded into a quantum amplitude-embedding circuit. After executing the circuit with 4096 shots, measured probabilities are rescaled to recover the original bin counts, which in turn reconstruct block-level intensities. Block sums are reconstructed, and the full image is reassembled, after which PSNR and MSE are evaluated.

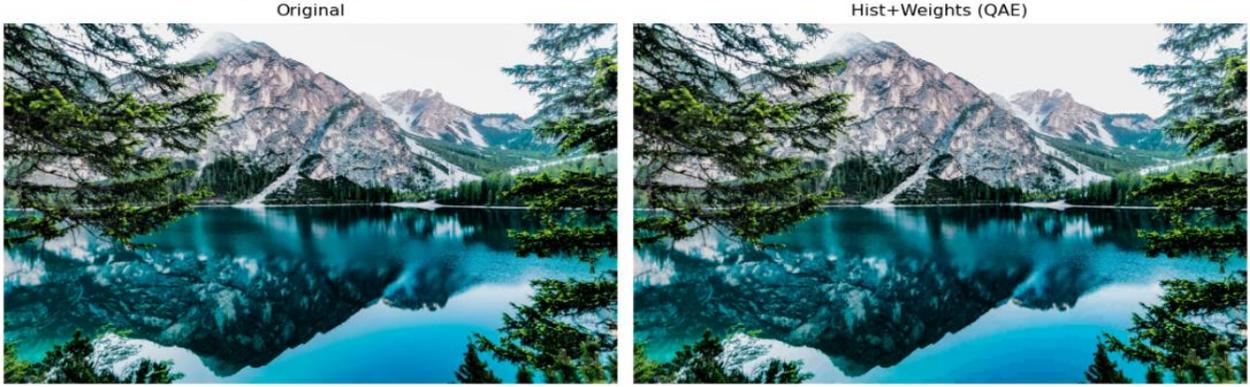

Figure 4: Visual comparison of original and reconstructed images using the proposed 5-qubit histogram-weighted encoding method on an actual IBM Quantum hardware using the ibm_sherbrooke for a random test image.

A bin size of 32 (i.e., 5 qubits) was selected for hardware execution to balance reconstruction fidelity and hardware constraints. This setting provides a mid-level histogram resolution with shallow circuit depth, making it feasible for noisy intermediate-scale quantum (NISQ) devices while still capturing sufficient image detail for meaningful reconstruction. The proposed approach has been executed on actual IBM Quantum hardware using Qiskit Runtime v2 through PennyLane integration. The execution was performed on a high-resolution test image (3403×5266×3) (≈1.79×10$^7$ pixels, ≈5.36×10$^7$ RGB values), with a 5-qubit histogram binning configuration. A visual comparison of the original and reconstructed outputs is shown in Figure 3. FRQI requires 25 position qubits plus one angle qubit (26 total) for grayscale image, NEQR uses 25 position qubits and 24 value qubits (49 total), and NCQI employs 25 plus two complex-amplitude qubits (27 total). By contrast, the histogram approach needs only $\lceil \log_2 B \rceil$=5 qubits in total for bin size of 32 even for this high-resolution image. The proposed histogram-based amplitude embedding scheme not only matches but, in many cases, surpasses the minimal qubit requirements of FRQI and NCQI, while introducing several critical enhancements in scalability, precision, and practicality. Unlike prior methods that constrain inputs to square grids (e.g., $2^n \times 2^n$), the present approach supports arbitrary image dimensions and aspect ratios. Only local block-level (bixel) padding is needed to ensure alignment, offering greater flexibility for real-world image datasets. Rather than encoding raw pixel values into quantum amplitudes as in classical amplitude embedding, this method compacts structural information by summarizing local block intensities and embedding their global histogram into a quantum amplitude vector. This strategy eliminates the need for per-pixel or per-channel registers required in formats such as NEQR, while still preserving enough structural detail for high-quality image reconstruction.

Table 1: Qubit and capability comparison across quantum image encoding methods

| Method | Qubits Required | Color Support | Input Size Constraint |
|---|---|---|---|
| **FRQI [2]** | $\lceil \log_2 N \rceil + 1$ | Grayscale | Square images only (N = $2^{2n}$ pixels) |
| **NEQR [4]** | $\lceil \log_2 N \rceil + l$, l = bit depth (typically 8) | Grayscale | Square images only |
| **NCQI [5]** | $\lceil \log_2 N \rceil + 2$ | RGB | Square images only |
| **Proposed** | $\lceil \log_2(B) \rceil$ | RGB (via bixel histogram) | Arbitrary image sizes and aspect ratios |

To validate reconstruction efficiency in practical terms, the quantum-reconstructed RGB images were saved as lossless PNG files and compared against their classical originals. File sizes were found to be nearly identical, confirming that histogram-based quantum encoding introduces no perceivable degradation in visual quality or storage efficiency, despite the radical compression at the quantum level. Together, these features establish the proposed method as a robust, scalable, and resource-efficient quantum image compression strategy poised for deployment in both simulation and future hardware pipelines.

## 5. CONCLUSION & FUTURE RESEARCH DIRECTIONS

A quantum image compression framework has been introduced that departs from conventional pixel-wise amplitude embedding in favor of a histogram-based encoding strategy, which captures bixel-level intensity distributions. By embedding a global histogram of local block sums into an amplitude vector, the method compresses arbitrarily large RGB images using only $\lceil \log_2 B \rceil$ qubits, where B is the number of histogram bins. This results in a minimal quantum footprint without the need for additional value registers or enforced square image shapes, distinguishing it significantly from earlier formats such as FRQI, NEQR, or NCQI.

Empirical evaluations on ibm_sherbrooke, using 4,096 shots per circuit, confirm that the proposed circuit is not just simulation-ready but also executable on current NISQ hardware. Hardware results demonstrated that with only 5 qubits (32 bins), the circuit achieved ~38 dB PSNR, MSE around 0.0025, and execution time under 0.4 seconds. Increasing to 6 qubits (64 bins) improved PSNR to ~65 dB and lowered MSE below 0.001, while 7 qubits (128 bins) achieved ~100 dB PSNR, MSE under 0.0001, and maintained runtime below 0.5 seconds. These results confirm that the amplitude-only circuit delivers high-fidelity image reconstruction with extremely low qubit and time cost, making it highly viable for current quantum processors. Looking ahead, several research avenues remain open:

- Adaptive Histogram Binning: Dynamically optimizing bin distributions may further improve encoding precision or reduce required qubit depth.
- Noise-Resilient Implementation: While initial hardware tests show promise, deeper exploration under realistic noise models and larger images is necessary to assess robustness and explore mitigation techniques.
- Hybrid Quantum-Classical Schemes: Pairing quantum histogram embeddings with classical lightweight neural decoders could enable real-time reconstruction from limited quantum state measurements.
- Application-Specific Extensions: Adapting this approach to domain-specific imagery (e.g., satellite, hyperspectral, or medical) could lead to more efficient encodings by exploiting natural redundancy or sparsity.

## AUTHORS' BACKGROUND

| Your Name | Title* | Research Field | Personal website |
| --- | --- | --- | --- |
| Sahil Tomar | Scientist | Quantum Machine Learning | https://www.linkedin.com/in/tomarsahil |
| Sandeep Kumar | Senior Scientist | Quantum Machine Learning | https://www.researchgate.net/profile/Sandeep-Kumar-405?ev=hdr_xprf |